\documentclass{pnastwo}

\usepackage{graphicx}

\usepackage{amssymb,amsfonts,amsmath}

\newcommand{\eq}[1]{\begin{align} #1 \end{align}}

\contributor{Submitted to Proceedings
of the National Academy of Sciences of the United States of America}
\url{www.pnas.org/cgi/doi/10.1073/pnas.0709640104}
\copyrightyear{2008}
\issuedate{Issue Date}
\volume{Volume}
\issuenumber{Issue Number}

\usepackage{color}

\usepackage{txfonts} 

\begin{document}


\title{Tribonucleation of bubbles}





\author{Sander Wildeman\affil{1}{University of Twente}, Henri Lhuissier\affil{1}{}, Chao Sun\affil{1}{}, Detlef Lohse\affil{1}{}, \and Andrea Prosperetti\affil{1}{}}
\contributor{To be submitted to Proceedings of the National Academy of Sciences
of the United States of America}

\maketitle

\begin{article}

\begin{abstract}
We report on the nucleation of bubbles on solids that are gently rubbed against each other in a liquid. The phenomenon is found to depend strongly on the material and roughness of the solid surfaces. For a given surface, temperature and gas content, a trail of growing bubbles is observed if the rubbing force and velocity exceed a certain threshold. Direct observation through a transparent solid shows that each bubble in the trail results from the early coalescence of several microscopic bubbles, themselves detaching from microscopic gas pockets forming between the solids. From a detailed study of the wear tracks, with Atomic Force and Scanning Electron Microscopy imaging, we conclude that these microscopic gas pockets originate from a local fracturing of the surface asperities, possibly enhanced by chemical reactions at the freshly created surfaces. Our findings will be useful for either preventing undesired bubble formation, or, on the contrary, for `writing with bubbles', i.e. creating controlled patterns of microscopic bubbles.
\end{abstract}




\section{Introduction}
	Elementary considerations show that a bubble will spontaneously disappear unless its radius $r$ is larger than a critical value $r_c = 2 \gamma/\Delta P$, where $\gamma$ is the surface tension of the liquid and $\Delta P$ is the difference between the pressure of the bubble contents and the surrounding liquid \cite{brennen1995}. Only bubbles larger than $r_c$ can persist and grow by gas diffusion and liquid evaporation. The classical kinetic theory of nucleation \cite{skripov1974} shows that, for water, the \emph{spontaneous} formation of critical bubbles requires either superheats of 212\,$^\circ$C or negative pressures (i.e. tensions) of 140 MPa. Recent experiments have come close to the quantitative verification of these predictions \cite{mekki12, zheng1991}, but only at the cost of a great deal of sophistication and ingenuity. It must therefore be concluded that a different mechanism is responsible for the exceedingly commonplace occurrence of bubbles.

	The seed for the currently accepted explanation was planted by Gernez \cite{gernez1867} who, in 1867, hypothesized that bubbles start from a pre-existing gaseous nucleus lodged in solid impurities or the walls of the container. An explanation for the stability of these heterogeneous nuclei was later supplied by Harvey et al. \cite{harvey1944} who pointed out that the curvature induced by contact with a hydrophobic solid surface would be able to stabilize a gas pocket even in an under-saturated liquid. This `crevice model' of bubble nucleation explains a large number of observations and has been applied to the development of so-called enhanced boiling surfaces \cite{kotthoff2006, webb2004}. Gas bubbles can be further stabilized by the formation of organic skins at their surface \cite{fox_herzfeld1954, yount1978}.

In spite of these advances, the nucleation phenomenon still exhibits obscure facets, one of which -- tribonucleation -- is studied in this paper. It has been known for at least half a century that, as noticed by Hayward in 1967 \cite{hayward1967}, ``extremely gentle rubbing'' of two solid objects inside a liquid under tension, which is otherwise stable against most forms of mechanical action (e.g. knocking on the container wall or stirring), readily induces nucleation. This tribonucleation is often cited as a plausible source of the microbubbles found in the limbs of humans and animals after physical exercise \cite{mcdonough1984, wilbur2010}. Campbell \cite{campbell1968} and Ikels \cite{ikels1970} attributed the nucleation observed in these conditions to the pressure drop induced by the viscous flow in the space between two separating solid surfaces. Indeed, in highly viscous liquids bubble formation compatible with this picture has been reported \cite{ashmore2005, prokunin2004, chen1991}. However, this explanation cannot easily account for the nucleation observed in low viscosity fluids like water and ethanol, because in many cases the theoretical gap between the solids would have to be smaller than the surface roughness. More strikingly, it cannot account for the key observation by Hayward that bubbles do not nucleate in the case of a rolling motion, but only in the case of a sliding motion between the solids \cite{hayward1967}, although for the same force and velocity the pressure drop is expected to be twice as large for rolling than for sliding \cite{ONeil67}. Another instance of bubble nucleation upon solid-solid contact in a low viscosity liquid was reported by Theofanous et al. \cite{theofanous1978}. These authors were able to reliably nucleate single bubbles by gently bringing into contact two stainless steel wires in Freon superheated by up to 60\,$^\circ$C.

In this paper we present experiments in which we rub a bead against a wafer submerged in a low viscosity liquid. We vary the rubbing force and velocity, the temperature, and the materials of the solids. Our approach is to combine macroscopic observations, revealing a threshold for the rubbing induced nucleation, with microscopic observations at the smallest scales of the problem: that of the apparent (Hertz) contact between the solids and that of the roughness tips where the actual contact is realized.

\section{Preliminary findings}

\begin{figure}[b]
	\begin{center}\includegraphics[width=\columnwidth]{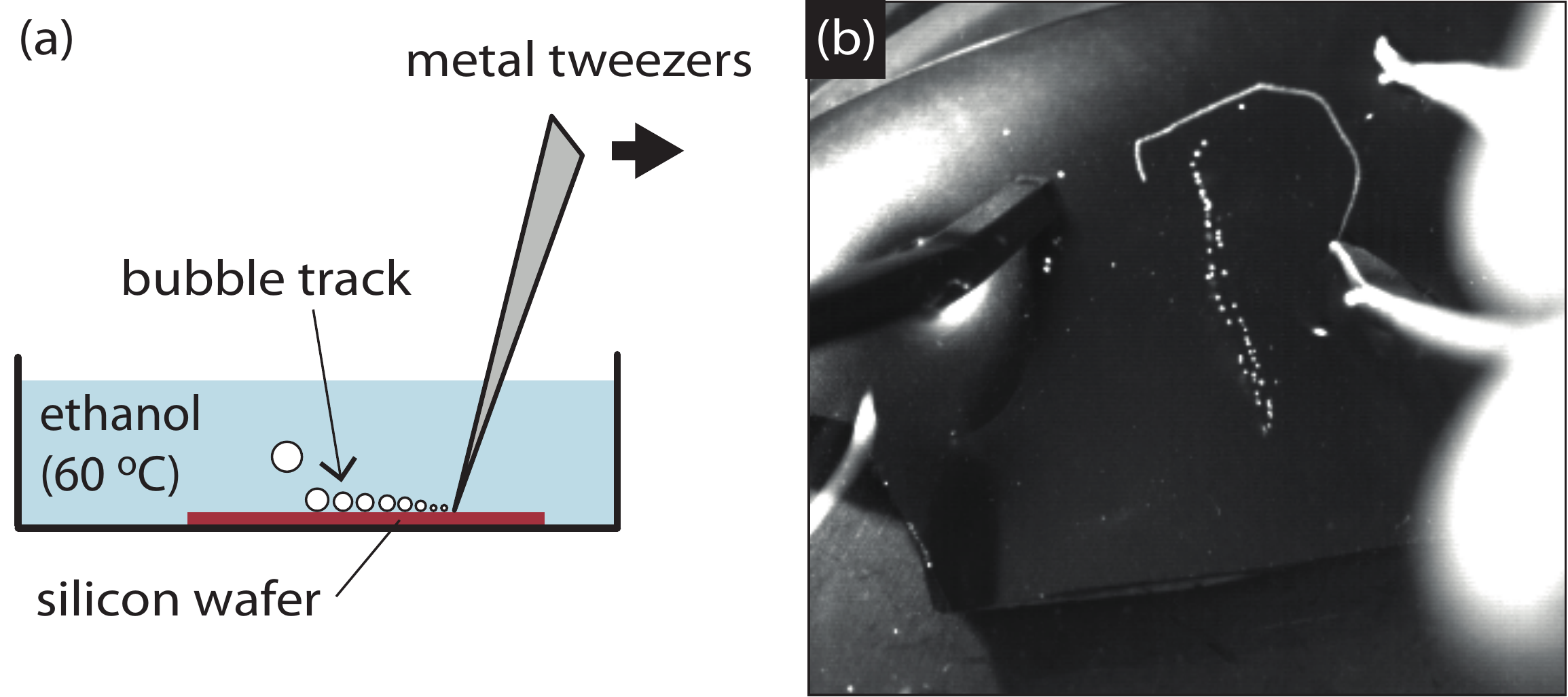}\end{center}
	\caption{(a) Schematic of the rubbing experiment. (b) Photograph of one of the authors writing a `P' with a trail of slowly growing bubbles by gently rubbing the tip of a pair of metal tweezers over a piece of unpolished silicon wafer submerged in ethanol. \label{fig:intro}}
\end{figure}

	This study was prompted by a recent observation in one of our experiments with heated liquids: bubbles are formed when the tip of a pair of stainless steel tweezers is gently rubbed against a submerged piece of unpolished silicon wafer (see figure \ref{fig:intro}). When the temperature of the liquid is below the boiling point, the bubbles appear as a trail behind the tweezers, where they slowly grow until they detach. After the detachment, no new bubbles are formed, indicating that the rubbing does not create permanent nucleation sites.

	This tribonucleation phenomenon occurred in all fluids we have tried so far (ethanol, water, acetone, pentane and perfluorohexane), whether polar or not, and whether they wet the solids or not. Conversely, it depends strongly on the material of the wafer being rubbed. Bubbles appear on silicon and aluminium under mild rubbing conditions, but remain absent on copper, glass and sapphire even upon vigorous rubbing.\footnote{Bubbles did appear at high loads on glass and sapphire with the tweezers, but not when the tweezers were replaced by a glass or sapphire bead. We think that this is because the tweezers material itself promotes the tribonucleation to some extent.}

\begin{figure*}[t]
	\begin{center}\includegraphics[width=\textwidth]{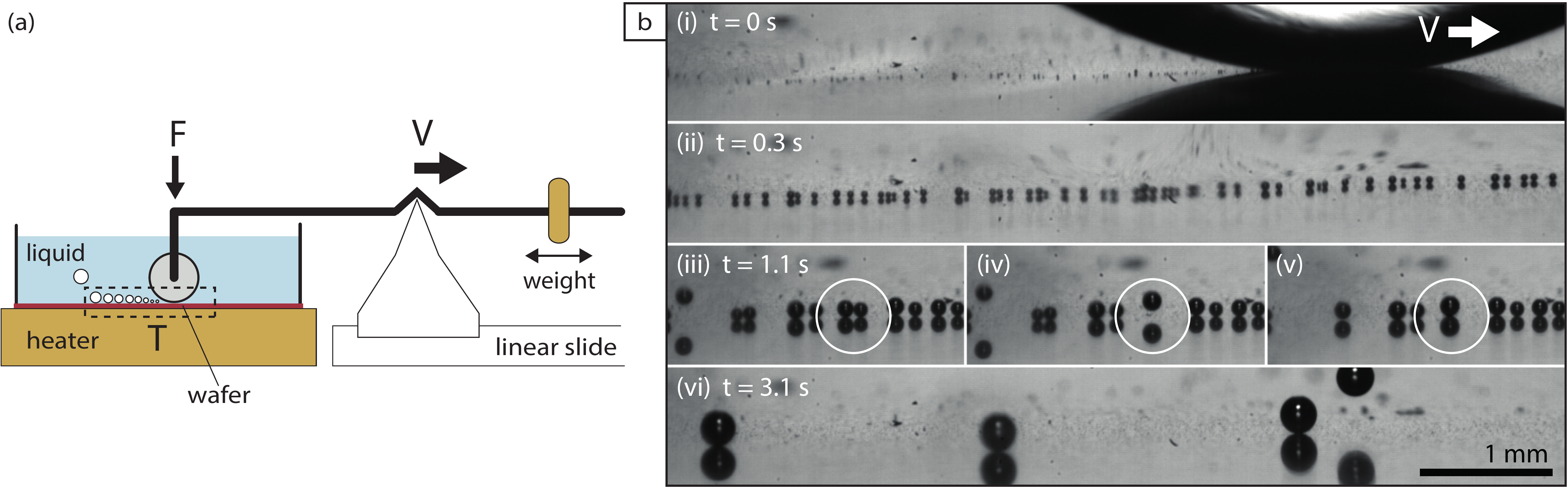}\end{center} 
	\caption{(a) Experimental setup. It consists of a lever to control the normal force on the bead, a linear slide (connected to a linear motor) to control the rubbing velocity and a heater to set the temperature at the bottom of the liquid cell. The wafer surface can be closely monitored from the side (dashed rectangle) or from the top with a long-distance microscope. (b) Side snapshots of the experiment. Due to the backlighting conditions, bubbles (and their reflection in the silicon wafer) appear as black disks against a lighter background. (i) A smooth sapphire bead, submerged in ethanol, is rubbed from left to right against an unpolished silicon wafer at 70\,$^\circ$C. (ii) The bubble trail left behind the bead slowly grows by gas diffusion. (iii--v) When two bubbles touch (see white circle), they merge, jump up and then slowly settle down again due to the temperature gradient near the surface (Marangoni effect). (vi) Eventually, buoyancy overcomes the downward Marangoni force and the bubbles rise to the free surface. \label{fig:setup}}
\end{figure*}

\section{Force-velocity dependence}

	In order to obtain quantitative information about the rubbing conditions for bubble formation, we used the setup sketched in figure \ref{fig:setup}(a). A smooth sapphire bead (average roughness $\mathcal{R}_a<0.025\,\muup$m from Ceratec; radius $R=4$\,mm) attached to a movable lever arm, replaces the tweezers. A heater was placed underneath the wafer being rubbed. This allowed us to precisely control the nominal geometry of the contact, the normal force $F$ applied to the bead, the rubbing velocity $V$ and the temperature $T$ of the submerged surfaces. Ethanol (99.8$\%$ from Assink Chemie; boiling point 78\,$^\circ$C) was used as the liquid.

	Figure \ref{fig:setup}(b) shows a typical experiment, observed from the side through a long-distance microscope. At $t = 0$\,s (i) the bead rubs against the \emph{unpolished} side of a silicon wafer held at a temperature of 70\,$^\circ$C. Subsequently (ii--vi), small bubbles (black spots) appear behind the bead and slowly grow by gas diffusion. A theoretical estimate of the relevant timescales of this growth is provided in the Supplementary Material. Although ethanol wets the silicon wafer, the bubbles do not immediately detach. They are pulled down by a `Marangoni force' induced by the temperature gradient close to the wafer surface \cite{young1959}. Snapshots (iii--vi) show how two bubbles in the row merge, jump up and then settle down again. The jumping is driven by a release of surface energy during merger, as described for droplets in reference \cite{boreyko2009}.  As the bubbles grow bigger, the upward buoyancy force eventually overcomes the downward Marangoni force and they rise to the free surface (vi).

\begin{figure}
\begin{center}\includegraphics[width=.9\columnwidth]{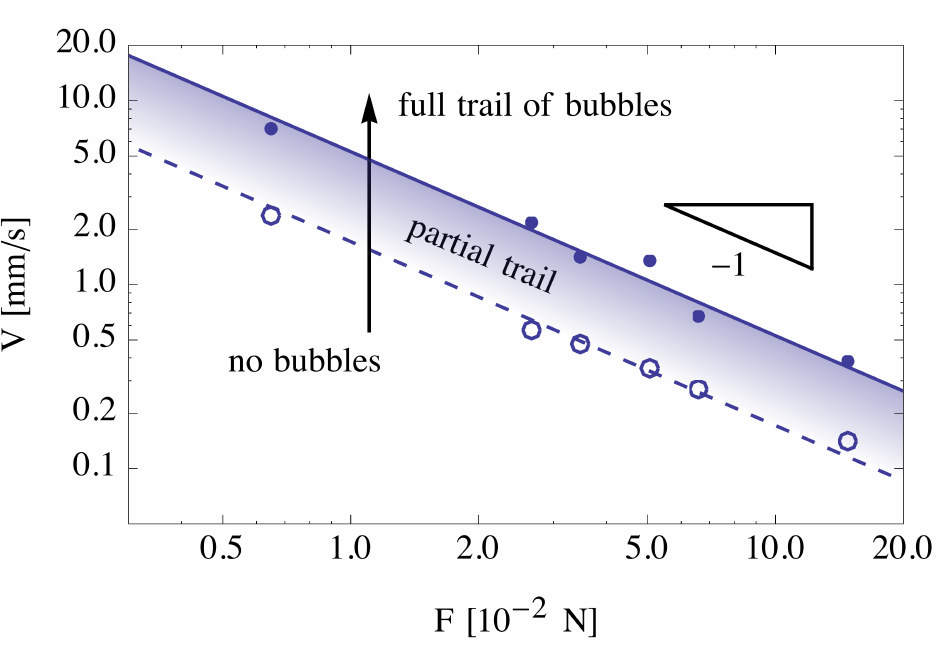} \end{center} 
	\caption{Threshold velocity for the formation of a trail of bubbles as a function of the normal force, for a sapphire bead and an unpolished silicon wafer. The wafer temperature is 70\,$^\circ$C and the fitted lines correspond to $FV = \text{const}$. \label{fig:sithreshold}}
\end{figure}

	The experiment was repeated for different rubbing velocities and loads (figure \ref{fig:sithreshold}). In each experiment, we fixed the normal force on the bead and then increased the velocity step by step, while monitoring the bubble trail behind the bead. We distinguished between a `full trail of bubbles', a `partial trail of bubbles' and `no bubbles'. The data shows that the higher the load, the lower the rubbing velocities required to generate bubbles. As indicated by the lines in figure\,\ref{fig:sithreshold}, the thresholds we measured are well described by 
\eq{
FV &= \text{const},
} 
with $\text{const} = 17\,\muup$W and $53\,\muup$W for the lower and upper thresholds, respectively.

\section{Influence of the material and surface roughness}

	The fact that no sharp transition exists from no bubbles to a full trail suggests that bubble formation depends, somehow, on the varying conditions along the rubbing track. Indeed, when the \emph{polished} side of the silicon wafer was rubbed, the generation of bubbles became significantly harder and less regular. Moreover, when the bead was continuously rubbed back and forth over the same track on the unpolished wafer, bubble formation stopped after typically 10 to 20 strokes, suggesting that rubbing locally changes the surface. If, subsequently, the bead was slightly displaced from the deactivated track during the rubbing, bubbles formed again.

\begin{figure}
\begin{center}\includegraphics[width=.87\columnwidth]{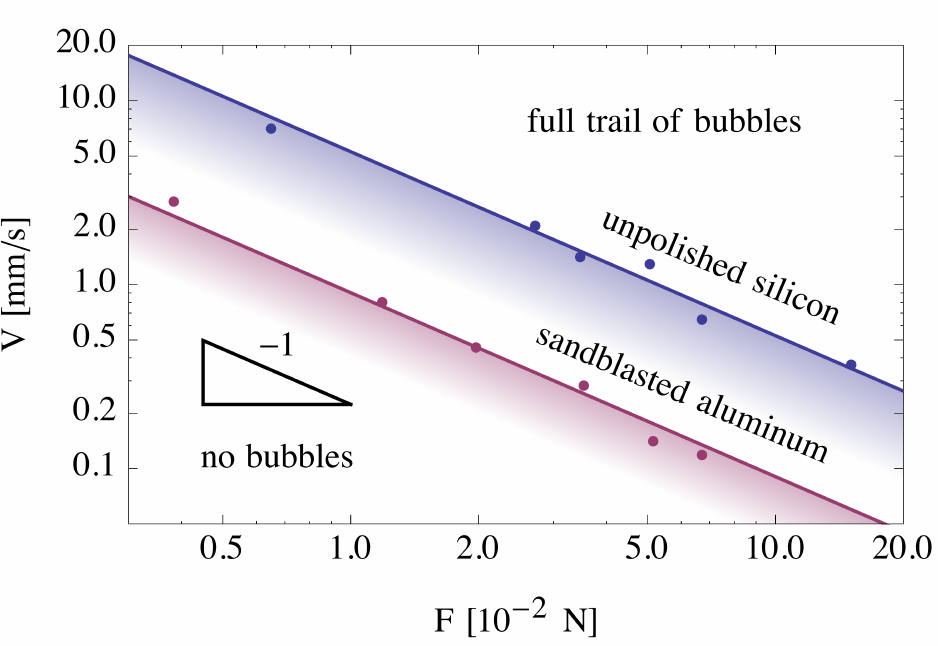}\end{center} 
	\caption{Comparison between the force-velocity thresholds for unpolished silicon and sandblasted aluminum submerged in ethanol at 70\,$^\circ$C and having similar roughness $\mathcal{R}_a\sim 0.5\,\muup$m. The solid lines correspond to $FV = \text{const}$. In the shaded area we still observe a partial bubble trail, see figure \ref{fig:sithreshold}. \label{fig:sivsal}}
\end{figure}

	Besides silicon, bubbles are also readily formed on aluminium. In figure \ref{fig:sivsal} we compare the tribonucleation threshold for aluminium with that for silicon. To give aluminium a macroscopic roughness similar to that of the silicon wafer ($\mathcal{R}_a\sim0.5\,\muup$m), the surface was sandblasted with a fine grain before the experiment. The threshold for aluminium turns out to be significantly lower than for silicon ($9\,\muup$W as compared to $53\,\muup$W). Moreover, in contrast to the quick deactivation of the rubbing tracks on silicon, the tracks on aluminium kept on generating bubbles even after more than 1000 strokes (the largest value we tried), although after typically 20 strokes a polished wear track became clearly visible on the roughened aluminium. Lastly, when the sandblasted surface was replaced by a smooth layer of aluminium (vapor-deposited on a glass slide, $\mathcal{R}_a\sim2\,$nm), no bubbles appeared during the first rubbing stroke, but did appear in subsequent passes over the same spot, hinting that the steady state wear track on aluminium is not smooth and promotes nucleation (as will be further discussed below).

\section{Influence of the temperature}

\begin{figure}[t]
\begin{center}\includegraphics[width=.8\columnwidth]{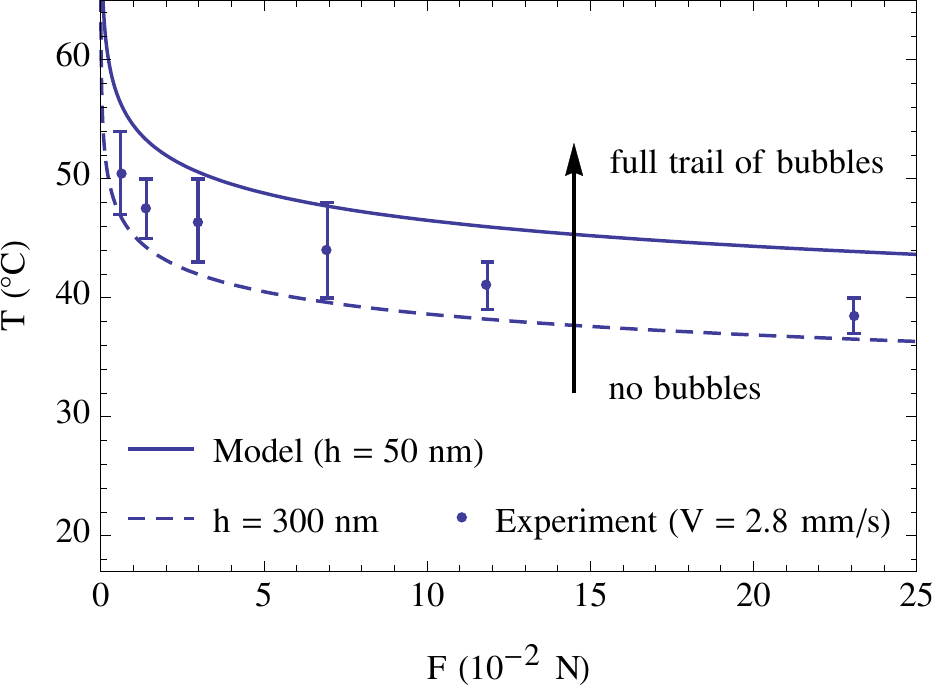}\end{center} 
	\caption{Threshold temperature for bubble formation as a function of the normal force, for a smooth sapphire bead rubbed against a polished aluminium surface with velocity $V=2.8$\,mm\,s$^{-1}$. The dots represent the experimental data, and the lines represent equation \eqref{eq:fc}. The single fitting parameter $h$ can be interpreted as a measure of the typical roughness height in the wear track (see text). \label{fig:Tthresh}}
\end{figure}

	In all the experiments described so far, the temperature was kept unchanged at 70\,$^\circ$C (about 8 degrees below the boiling point of ethanol). In order to determine if and how temperature affects the generation of bubbles, we did experiments in which we ramped the temperature from 25 to 70\,$^\circ$C, while continuously rubbing back and forth over the same track on a polished aluminium wafer. We choose aluminium because, on it, tracks do not deactivate but keep forming bubbles as long as the force-velocity threshold is overcome, as reported above. We fixed the velocity at 2.8 mm\,s$^{-1}$ (which at 70\,$^\circ$C is enough to generate bubbles at very low loads) and varied the normal force between each temperature ramp. The results are shown in figure \ref{fig:Tthresh}. They reveal that the lower the temperature, the higher the load required to generate bubbles. As with the force-velocity threshold (figure \ref{fig:sithreshold}), there is a finite transition region from no bubbles to a full trail. 

\section{Origin of the threshold for bubble trail formation}

\begin{figure}[t]
\begin{center}\includegraphics[width=.95\columnwidth]{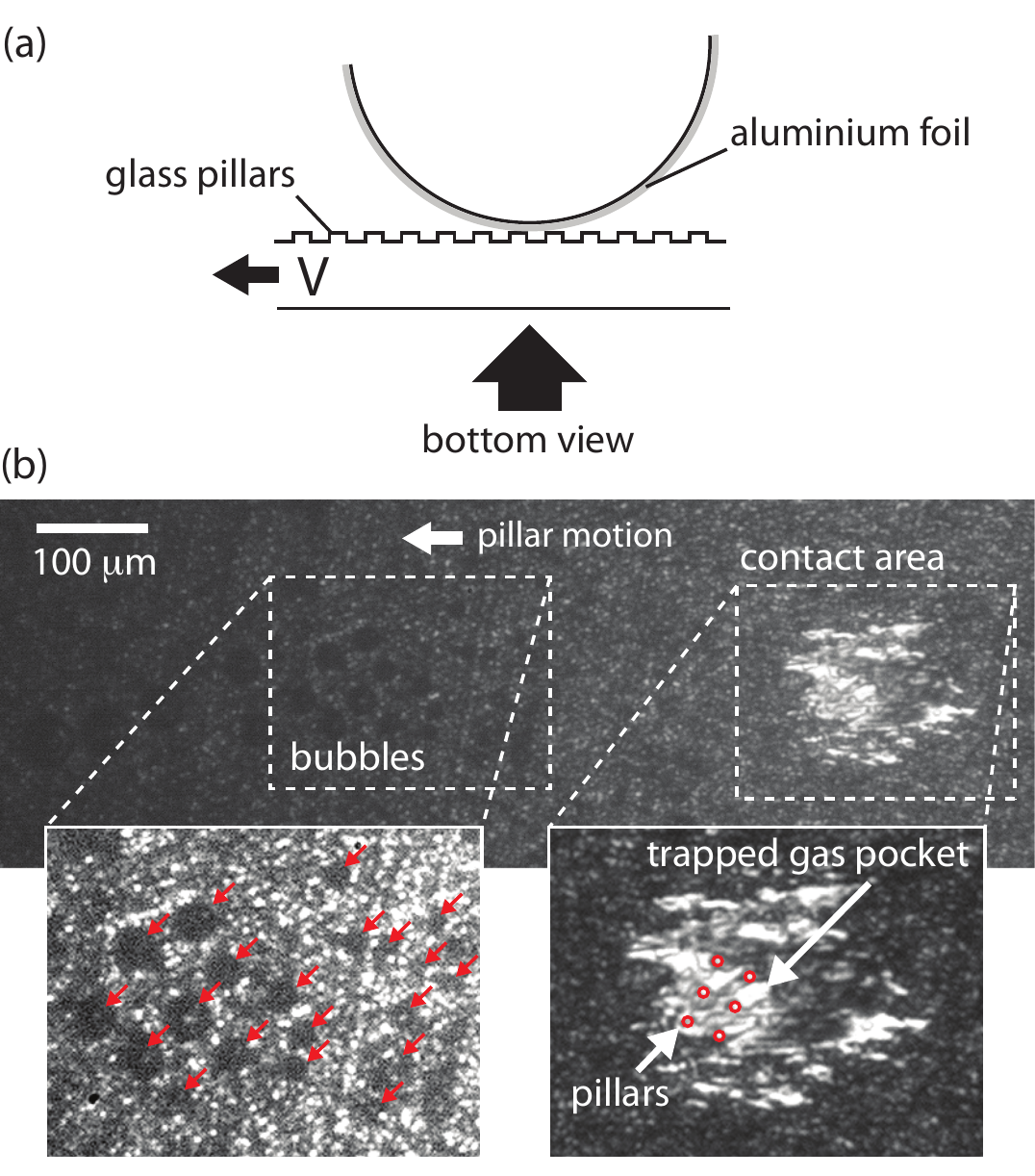}\end{center}
	\caption{(a) Schematic of the setup: a transparent glass surface covered with micro-pillars is rubbed against a bead covered with aluminium foil ($F = 7 \times 10^{-2}$\,N, $V = 4.7$\,mm\,s$^{-1}$). (b) Bottom view of the experiment through a microscope. Gas pockets are trapped between the two solids in the contact area and microscopic gas bubbles (red arrows) are observed downstream. The spacing between the pillars, their diameter and their height are 10\,$\muup$m, 9\,$\muup$m and 0.3\,$\muup$m, respectively. \label{fig:bottomview}}
\end{figure}

We envision the trail formation to depend on two steps: (1) the inception of bubble nuclei in the contact area and (2) the subsequent growth of these nuclei after the contact area has moved. To see how these steps determine the observed thresholds (figure \ref{fig:sivsal} and \ref{fig:Tthresh}), we set up the experiment shown in figure \ref{fig:bottomview}(a). A piece of aluminium foil was tightly wrapped around the sapphire bead, which was then rubbed against a glass substrate submerged in ethanol. This allows for a direct observation of the contact area through the glass. To enhance the wear of the foil by the substrate, the latter was equipped with protrusions in the form of micro-pillars.

	The direct observation of the contact area provided some crucial insights. First, as shown in figure \ref{fig:bottomview}(b), rubbing can trigger the nucleation of bubbles on aluminium even at room temperature. During the rubbing, gas continuously comes out of solution and collects in microscopic gas pockets trapped between the two solids. As the substrate moves on, these pockets are ejected in the form of small bubbles. At room temperature, these bubbles do not grow, but dissolve as soon as they reach the bulk of the liquid. Similarly, the gas pockets trapped in the contact area slowly dissolve when the substrate motion is stopped. Second, at temperatures for which the pressure inside the bubble is large enough to make them persist, the microscopic bubbles merge, resulting in a regular trail (figure \ref{fig:setup}b). This indicates that in figure \ref{fig:Tthresh} it is the supersaturation condition (and not the bubble inception) which dictates the trail formation.

\begin{figure}[t]
\begin{center}\includegraphics[width=.8\columnwidth]{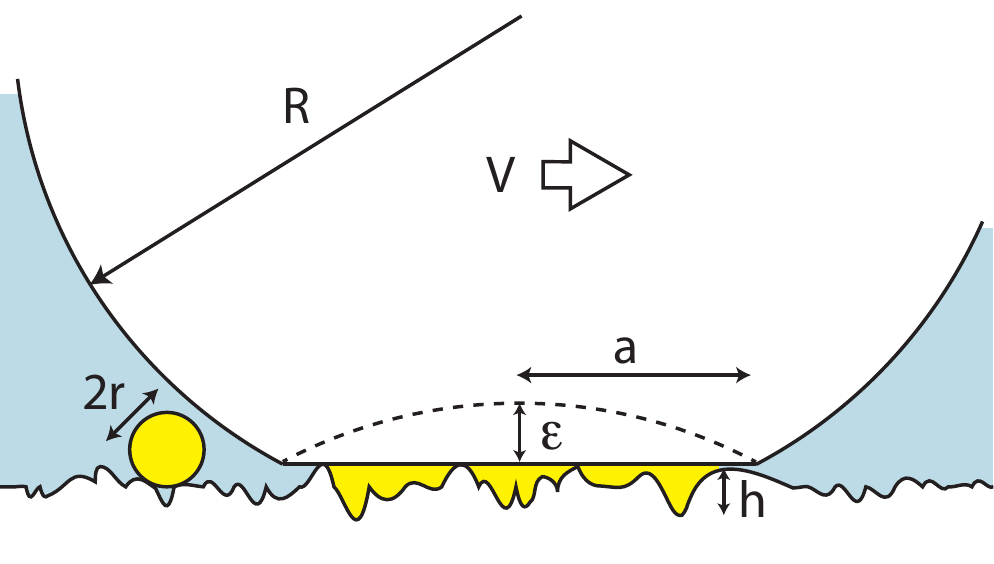}\end{center}
	\caption{Geometry used for the model. Gas (yellow) completely fills the gap in the apparent contact area and collects in a single bubble with radius $r$. For clarity the bead is shown indented here, while in reality the indentation is essentially concentrated on the aluminium surface which is much softer than the sapphire bead. \label{fig:hertz}}
\end{figure}

We can quantify this idea with a model in which the size of the microscopic bubbles is set by the space available in the contact area (see figure \ref{fig:hertz}). First we use Hertz's contact theory to estimate the radius $a$ of the apparent contact area
\eq{
	a \sim \sqrt{\epsilon R},
	\label{eq:hertz}
}
where the indentation depth $\epsilon$ is related to the normal force $F$ as \cite{landau1970}
\eq{
	\epsilon &\sim \left(\frac 34 \frac{F}{E^* R^{1/2}}\right)^{2/3}.
}
Here $E^*$ is the effective elastic modulus of the particular bead-substrate combination, which is dominated by the softer of the two. Combining the contact radius $a$ with a typical roughness height $h$ gives the volume available for all the gas pockets
\eq{
	\Omega \sim \pi a^2 h.
}
If we suppose that all this gas ends up in a single bubble of radius $r = (3\Omega/4\pi)^{1/3}$, then, as is mentioned in the introduction and elaborated on in the Supplementary Material, this unstable bubble will grow in the bulk if its radius is larger than
\eq{
	r_c &= \left(\frac{3\Omega_c}{4\pi}\right)^{1/3} \sim \frac{2\gamma}{\Delta P}.
	\label{eq:rc}
}
Prior to being heated, the liquid used in the experiments was equilibrated for a long time with air (i.e. gas + vapor) at a temperature of $T_0 = 20\,^\circ$C; under a total pressure $P_{atm} = P_g + P_v(T_0) = 1\,\text{bar}$. During the heating to a temperature $T$, the gas content of the liquid, i.e. $P_g \simeq P_{atm} - P_v(T_0)$, did not change appreciably (since the relevant gas diffusion timescale is much longer than that of the heating, and the gas solubility only changes by 10\% over the temperature range). The excess pressure $\Delta P$ in the heated liquid therefore comes down to the increase in the vapor pressure from $T_0$ to $T$, that is, $\Delta P \simeq P_v(T) - P_v(T_0) = \Delta P_v(T)$. Combining equations \eqref{eq:hertz} -- \eqref{eq:rc} yields an expression for the critical force as a function of temperature
\eq{
	F_c(T) &\sim \frac{4E^*}{3 R}\left(\frac{4}{3h}\right)^{3/2} \left(\frac{2\gamma}{\Delta P_v(T)}\right)^{9/2},
	\label{eq:fc}
}
with a pre-factor $\sim 1$.

	Equation \eqref{eq:fc} is plotted in figure \ref{fig:Tthresh} for $h = 50$ and $300$\,nm, using the surface tension $\gamma = 0.02$\,N\,m$^{-1}$ of ethanol, the elastic modulus $E^* = 70$\,GPa of aluminium, and an empirical relation for the ethanol vapor pressure \cite{ambrose1970}. The roughness parameters of 50 and 300\,nm correspond, in order of magnitude, to the large and small scale roughnesses we measured by Atomic Force Microscopy (AFM) on the steady state wear track on aluminium. The experimental data in figure \ref{fig:Tthresh} therefore seems to be consistent with our model. One crucial question however remains: what controls the \emph{nucleation} of these gas pockets?

\section{Microscopic mechanism for gas pocket formation}

\begin{figure}[t]
\begin{center}\includegraphics[width=\columnwidth]{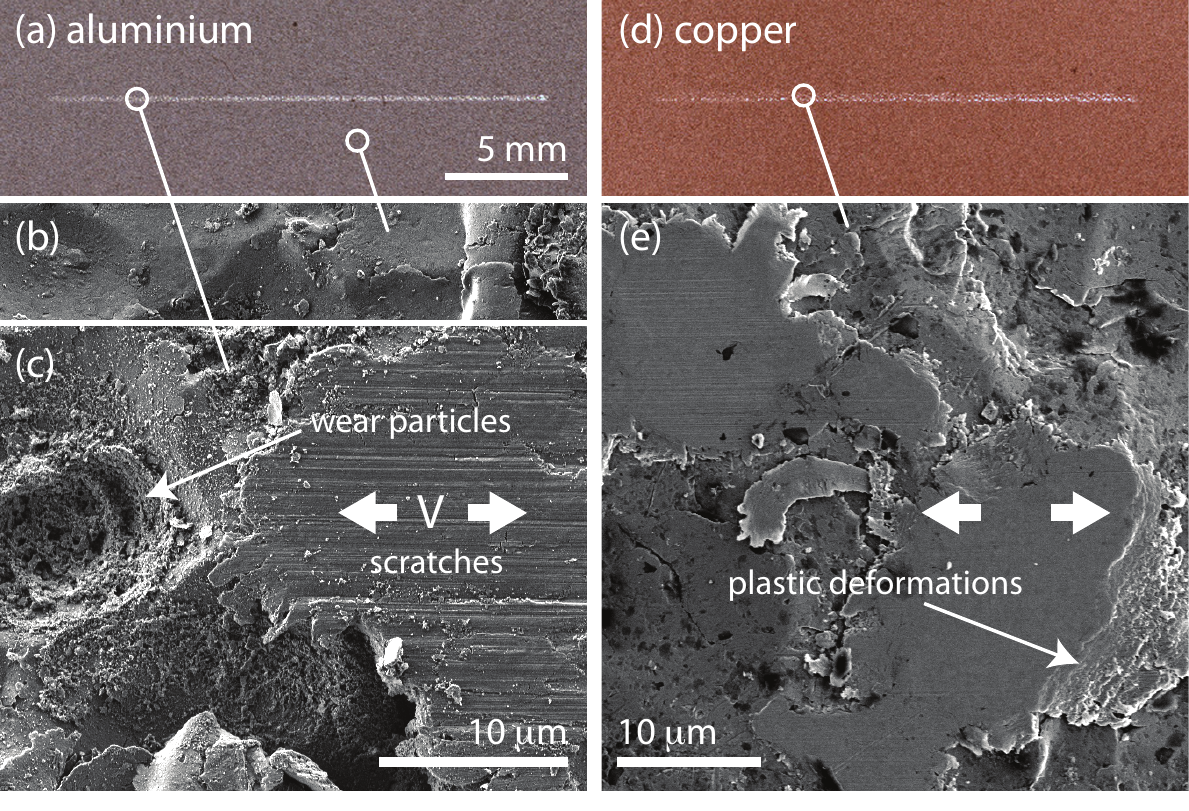}\end{center}
	\caption{Wear tracks formed by rubbing a sapphire bead 50 times back and forth against (a,b,c) sandblasted aluminium and (d,e) sandblasted copper. Rubbing parameters: $T = 70\,^\circ$C,  $F = 7\times10^{-2}$\,N and $V=1$\,mm\,s$^{-1}$. (a,d) The wear tracks are clearly visible under racking lighting and look very similar to the naked eye. (c,e) SEM imaging however reveals important differences at the scale of a single asperity. (c) On aluminium, the scratches on top of the flattened asperities are relatively deep ($\mathcal{R}_a\sim 50$\,nm from AFM measurements) and many small wear particles are collected in the troughs around the asperities (as for a brittle material). (e) On copper, the tops of the asperities are much smoother ($\mathcal{R}_a\sim 5$\,nm), they seem to be plastically squeezed, and no wear particles are observed (as for a ductile material). \label{fig:weartracks}}
\end{figure}

	As mentioned in the `Preliminary findings', tribonucleation is observed on aluminium, but not on copper. This prompted us to closely analyze the wear tracks left on each surface. Figure \ref{fig:weartracks} shows photographs and detailed Scanning Electron Microscopy (SEM) recordings of wear tracks on aluminium and copper. Although the two tracks look very similar optically, they can be easily distinguished in the SEM images. Indeed, for both aluminium and copper, the asperities in the wear tracks got flattened by the rubbing bead. However, on aluminium this flattening seems to result from a continuous breaking off of small parts of the rough surface (as for a brittle material). The top surface of the flattened asperities has small scratches throughout and a lot of small wear particles are observed around, in the troughs. In contrast, for copper the top surface has a relatively low roughness, material is plastically squeezed out at the sides of the asperities, and no wear particles are observed around (as for a ductile material).

\begin{figure}[t]
\begin{center}\includegraphics[width=.9\columnwidth]{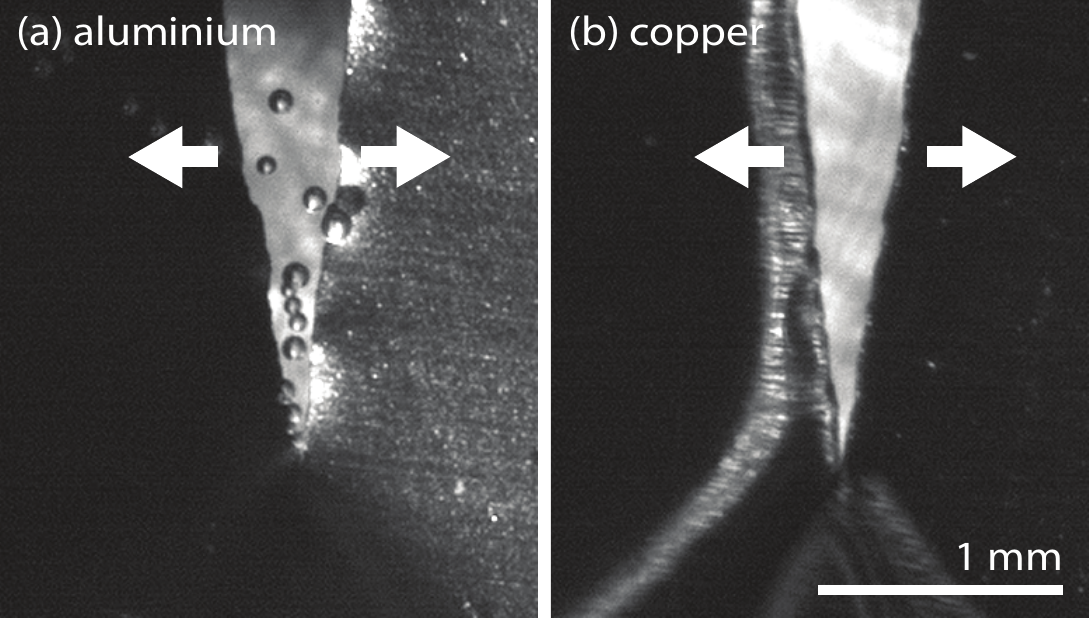}\end{center}
	\caption{Snapshots of the tearing of thin foils of (a) aluminium and (b) copper submerged in ethanol at around 78\,$^\circ$C. The foils are respectively 12.0\,$\muup$m and 12.5\,$\muup$m thick, and the two sides of the foil are torn apart with a velocity of about 9\,mm\,s$^{-1}$. For aluminium, bubbles form at the tip of the tear, while for copper no bubbles form. \label{fig:tear}}
\end{figure}

	The comparison of the wear tracks suggests that the fracturing of the surface is an essential ingredient for a material to provoke tribonucleation. To test whether it is the fracturing itself, or its products (i.e. the wear particles left on the track) which are responsible for the creation of gas pockets, we realized the experiment shown in figure \ref{fig:tear}. A piece of aluminium (soft Al99.5\%, Salamon's Metalen) or copper (Cu99.95\%, Salamon's Metalen) foil was immersed in hot ethanol and then torn apart at a constant velocity of about 9\,mm\,s$^{-1}$. Consistent with the rubbing experiments, bubbles did appear in the case of aluminium, but \emph{not} in the case of copper\footnote{Sometimes, a single bubble appeared in the very last stage, when the two ends of the copper foil completely separated at much higher velocity.}. Since wear particles play no role here, this experiment shows that the fracturing itself can generate gas nuclei.

\section{Discussion}
A possible scenario for the nucleation by fracturing is that the gap formed when a micro-crack opens fills with gas and vapor before the liquid can enter it. This embryo can then act as a nucleus for the formation of a visible bubble. Only in a brittle material this crack opening would be rapid enough.

While bulk aluminium is ductile, it has a thin ($\sim$\,nm) oxide layer on its surface, which might explain its brittle behavior. A similar layer exists on the surface of silicon. Note that this `passivation layer' forms because the bare materials readily react with any oxidizing molecules in their environment. In particular, in the presence of water this chemical activity causes the generation of hydrogen gas, which might be at the root of, or at least contribute to, the initial nucleation process. This idea is supported by experiments with degassed water, in which we still observed the formation of microscopic gas bubbles, which quickly dissolved after the rubbing had stopped. On the other hand we should stress that we could also create bubble trails in perfluorohexane (FC72), a liquid that should be inert in most circumstances. Also in the work by Theofanous et al. \cite{theofanous1978}, which involved polished stainless steel and Freon, chemical reactions are very unlikely.

The embryos formed by the mechanism of fracturing and, possibly, chemical reactions may not grow individually, but only if they merge with others before dissolving. The force-velocity threshold $FV$ = const. would then be a manifestation of the competition between generation, merging and dissolution and could be interpreted as a minimal frictional power input required for abundant local fracturing.

\section{Summary and conclusion}

We have found that bubbles are readily and reliably generated upon gentle rubbing of certain solid surfaces. An intriguing demonstration of the phenomenon is the ``writing'' example shown in figure \ref{fig:intro}. It is somewhat surprising that, in spite of its robustness and repeatability, this phenomenon has been the object of so little attention in the literature. 

We observed that bubbles can nucleate and form a trail on submerged solids under gentle rubbing conditions (normal force $F = 1 - 200$\,mN and relative velocity $V = 0.1-20$\,mm\,s$^{-1}$). At room temperature, small bubbles are observed to form and detach, but they dissolve as they move away from the contact area. As the temperature is increased, the bubbles persist and grow forming a trail. On silicon and aluminium, measurements in ethanol at 70 $^\circ$C and above indicate the existence of a threshold for the trail formation of the form $FV$ = const., with a constant 6 times larger for silicon than for aluminium.

Bubble formation strongly depends on the materials being rubbed. On silicon, tribonucleation stops after typically 20 strokes over the same spot, while on aluminium a steady state wear track forms from which bubbles keep appearing upon rubbing. Bubbles do not form on copper, although the wear tracks on copper and aluminium look very similar optically. SEM imaging shows that aluminium asperities are abraded by a fracturing, brittle-like mechanism, while copper asperities are flattened by plastic, ductile-like deformations. Additional experiments on the slow rupture of aluminium and copper foils indicate that fracturing alone (in the absence of wear) is sufficient to create gas nuclei.

The above observations evidence that trail formation by tribonucleation involves two steps: (1) bubble nucleation in the contact region and (2) subsequent growth of these nuclei in the bulk. Both steps need to be satisfied in order to see a trail. Our experiments show that fracturing is essential for the first step. We hypothesize that a void created by the rapid fracturing of the surface asperities, possibly in combination with chemical reactions forming gas at the freshly created surfaces, can explain the nucleation of bubbles at the low loads and velocities used in the experiments. When there is abundant local fracturing, the amount of gas that comes out of solution is limited by the space available between the asperities of the surfaces in contact, and it is this volume that sets the threshold for trail formation in this case.

We hope that the present exploratory work may motivate further studies to look into the fundamental mechanism(s) involved in tribonucleation, and to explain, for example, the emergent dependence of the phenomenon on the rubbing force and velocity.





\begin{acknowledgments}
We would like to thank Yanbo Xie for providing us with structured glass substrates, Gert-Wim Bruggert for his assistance with the design and implementation of the set up, Robin Berkelaar for help with scanning the wear tracks with an AFM and interpreting the images, Mark Smithers and Gerard Kip, associated with the MESA+ institute, for making high resolution SEM recordings for us, Erik de Vries from the Surface Technology \& Tribology group for useful discussion on the wear track images, Vincent Craig for emphasising the possible role of chemistry in the experiments and an anonymous referee for useful suggestions. This project was financed by an ERC Advanced Grant.
\end{acknowledgments}


\begin{thebibliography}{10}

\bibitem{brennen1995}
Brennen, C.~E.
\newblock (1995) {\em Cavitation and Bubble Dynamics}.
\newblock (Oxford University Press).

\bibitem{skripov1974}
Skripov, V.
\newblock (1974) {\em Metastable liquids}.
\newblock (J. Wiley).

\bibitem{mekki12}
El~Mekki~Azouzi, M, Ramboz, C, Lenain, J.-F,  \& Caupin, F.
\newblock (2012) {A coherent picture of water at extreme negative pressure}.
\newblock {\em Nature Phys.} {\bf 9}, 38--41.

\bibitem{zheng1991}
Zheng, Q, Durben, D.~J, Wolf, G.~H,  \& Angell, C.~A.
\newblock (1991) {Liquids at large negative pressures: Water at the homogeneous
  nucleation limit}.
\newblock {\em Science} {\bf 254}, 829--832.

\bibitem{gernez1867}
Gernez, M.
\newblock (1867) {On the disengagement of gases from their saturated
  solutions}.
\newblock {\em Phil. Mag.} {\bf 33}, 479--481.

\bibitem{harvey1944}
Harvey, E.~N, Whiteley, A.~H, McElroy, W.~D, Pease, D.~C,  \& Barnes, D.~K.
\newblock (1944) Bubble formation in animals.
\newblock {\em J. Cellular Comp. Physiol.} {\bf 24}, 1--34.

\bibitem{kotthoff2006}
Kotthoff, S, Gorenflo, D, Danger, E,  \& Luke, A.
\newblock (2006) {Heat transfer and bubble formation in pool boiling: Effect of
  basic surface modifications for heat transfer enhancement}.
\newblock {\em Int. J. Therm. Sci.} {\bf 45}, 217--236.

\bibitem{webb2004}
Webb, R.
\newblock (2004) {Odyssey of the enhanced boiling surface}.
\newblock {\em J. Heat Trans.-T. ASME} {\bf 126}, 1051--1059.

\bibitem{fox_herzfeld1954}
Fox, F.~E \& Herzfeld, K.~F.
\newblock (1954) Gas bubbles with organic skin as cavitation nuclei.
\newblock {\em J. Acoust. Soc. Am.} {\bf 26}, 984.

\bibitem{yount1978}
Yount, D.~E.
\newblock (1979) Skins of varying permeability: A stabilization mechanism for
  gas cavitation nuclei.
\newblock {\em J. Acoust. Soc. Am.} {\bf 65}, 1429.

\bibitem{hayward1967}
Hayward, A. T.~J.
\newblock (1967) Tribonucleation of bubbles.
\newblock {\em Brit. J. Appl. Phys.} {\bf 18}, 641.

\bibitem{mcdonough1984}
McDonough, P.~M \& Hemmingsen, E.~A.
\newblock (1984) Bubble formation in crabs induced by limb motions after
  decompression.
\newblock {\em J. Appl. Physiol.} {\bf 57}, 117--122.

\bibitem{wilbur2010}
Wilbur, J.~C, Phillips, S.~D, Donoghue, T.~G, Alvarenga, D.~L, Knaus, D.~A,
  Magari, P.~J,  \& Buckey, J.
\newblock (2010) Signals consistent with microbubbles detected in legs of
  normal human subjects after exercise.
\newblock {\em J. Appl. Physiol.} {\bf 108}, 240--244.

\bibitem{campbell1968}
Campbell, J.
\newblock (1968) The tribonucleation of bubbles.
\newblock {\em Brit. J. Appl. Phys.} {\bf 1}, 1085.

\bibitem{ikels1970}
Ikels, K.~G.
\newblock (1970) Production of gas bubbles in fluids by tribonucleation.
\newblock {\em J. Appl. Physiol.} {\bf 28}, 524.

\bibitem{ashmore2005}
Ashmore, J, del Pino, C, \& Mullin, T
\newblock (2005) Cavitation in a Lubrication Flow between a Moving Sphere and a Boundary.
\newblock {\em Phys. Rev. Lett.} {\bf 94}, 124501.

\bibitem{prokunin2004}
Prokunin, A.~N.
\newblock (2004) Microcavitation in the Slow Motion of a Solid Spherical Particle along a Wall in a Fluid.
\newblock {\em Fluid Dyn.} {\bf 39}, 110--118.

\bibitem{chen1991}
Chen, Y.~L, \& Israelachvili, J,
\newblock (1991) New Mechanism of Cavitation Damage.
\newblock {\em Science} {\bf 252}, 1157--1160.

\bibitem{ONeil67}
O'Neill, M.~E \& Stewartson, K.
\newblock (1967) {On the slow motion of a sphere parallel to a nearby plane
  wall}.
\newblock {\em J. Fluid Mech.} {\bf 27}, 705.

\bibitem{theofanous1978}
Theofanous, T.~G, Bohrer, T.~H, Chang, M.~C,  \& Patel, P.~D.
\newblock (1978) {Experiments and Universal Growth Relations for Vapor Bubbles
  With Microlayers}.
\newblock {\em J. Heat Trans.-T. ASME} {\bf 100}, 41--48.

\bibitem{young1959}
Young, N.~O, Goldstein, J,  \& Block, M.~J.
\newblock (1959) The motion of bubbles in a vertical temperature gradient.
\newblock {\em J. Fluid Mech.} {\bf 6}, 350.

\bibitem{boreyko2009}
Boreyko, J.~B \& Chen, C.
\newblock (2009) {Self-Propelled Dropwise Condensate on Superhydrophobic
  Surfaces}.
\newblock {\em Phys. Rev. Lett.} {\bf 103}, 184501.

\bibitem{landau1970}
Landau, L \& Lifshitz, E.
\newblock (1970) {\em Theory of elasticity}.
\newblock (Pergamon Press), 2nd edition.

\bibitem{ambrose1970}
Ambrose, D \& Sprake, C. H.~S.
\newblock (1970) Thermodynamic properties of organic oxygen compounds xxv.
  vapour pressures and normal boiling temperatures of aliphatic alcohols.
\newblock {\em J. Chem. Thermodyn} {\bf 2}, 631--645.

\end{thebibliography}





\bibliographystyle{pnas2009}

\end{article}








\end{document}